\documentclass[twocolumn,showpacs,preprintnumbers,amsmath,amssymb]{revtex4}
\usepackage{epsfig}
\usepackage{amsmath}

\begin{document}

\title{Communities in Neuronal Complex Networks Revealed by Activation 
Patterns}

\author{Luciano da Fontoura Costa}
\affiliation{Institute of Physics at S\~ao Carlos, University of
S\~ao Paulo, PO Box 369, S\~ao Carlos, S\~ao Paulo, 13560-970 Brazil}

\date{29th Jan 2008}

\begin{abstract}
Recently, it has been shown that the communities in neuronal networks
of the integrate-and-fire type can be identified by considering
patterns containing the beginning times for each cell to receive the
first non-zero activation.  The received activity was integrated in
order to facilitate the spiking of each neuron and to constrain the
activation inside the communities, but no time decay of such
activation was considered.  The present article shows that, by taking
into account exponential decays of the stored activation, it is
possible to identify the communities also in terms of the patterns of
activation along the initial steps of the transient dynamics.  The
potential of this method is illustrated with respect to complex
neuronal networks involving four communities, each of a different type
(Erd\H{o}s-R\'eny, Barab\'asi-Albert, Watts-Strogatz as well as a
simple geographical model).  Though the consideration of activation
decay has been found to enhance the communities separation, too
intense decays tend to yield less discrimination.
\end{abstract}

\maketitle

\vspace{0.5cm}
\emph{`Zora's secret lies in the way your gaze runs over patterns
following one another as in a musical score...'
(I. Calvino, Inivisible Cities)}

\section{Introduction} 

Neuronal networks (e.g.~\cite{Haykin:1998,Anderson:1995,Squire:2003})
and complex networks (e.g.~\cite{Albert_Barab:2002, Newman:2003,
Dorogov_Mendes:2002, Costa_surv:2007}) can be understood as sister
research areas.  However, as the latter is much younger (especially
regarding the developments from 1999), these two sisters have yet to
get fully acquainted one another.  Such a natural integration has
already begun (e.g.~\cite{Stauffer_Hopfield, Stauffer_Costa,
Costa_revneur:2005, Kim:2004, Timme:2006, Osipov:2007, Hasegawa:2004,
Hasegawa:2005, Park:2006}) and is poised to continue to the point that
these two areas become not only close relatives, but also best
friends.  This integration is particularly interesting for both
neuronal networks and complex networks because of the complementation
of the approaches which have been respectively adopted.  More
specifically, while neuronal networks have relied strongly on pattern
recognition and dynamical systems, complex networks have been strongly
focusing on structure, with a recent surge of interest on dynamics
(e.g.~\cite{Newman:2003, Boccaletti:2006}).  However, as special
emphasis has been placed on the important problem of linear
synchronization (e.g.~\cite{Boccaletti:2006}), few works have
addressed non-linear or transient dynamics
(e.g.~\cite{Costa_Ising:2007,Costa_superedges:2008}).  In complex
networks, emphasis has been placed on the modularity of the
connections or community structure (e.g.~\cite{Watts_Strogatz:1998,
Girvan:2002, Zhou:2003, Newman:2004, Radicchi:2004, Hopcroft:2004,
Latapy:2005, Guimera:2005, Bagrow:2005, Capocci:2005, Arenas:2008,
Costa_comm:2006}), which has important implications for both the
structure and dynamics of networks.  The integration between neuronal
networks and complex networks is henceforth referred to as
\emph{complex neuronal networks}, which has special importance for
non-linear dynamical systems underlain by structured and complex
connectivity.

\begin{figure}[htb]
  \vspace{0.3cm} 
  \begin{center}
  \includegraphics[width=0.9\linewidth]{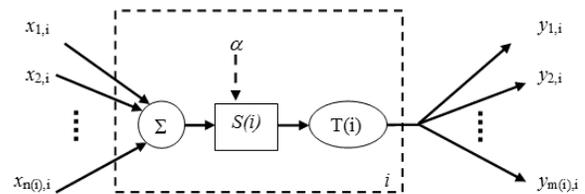} 
  \caption{The integrate-and-fire neuronal cell adopted in
               the previous works~\cite{Costa_nrn:2008,
               Costa_neucomm:2008} incorporates three stages: (i)
               integrating of input activations; (ii) memory
               of activation $S(i)$; and (iii) non-linear transfer
               function involving a threshold $T(i)$ (a hard limitter).
               While those previous works adopted full conservation
               of the activation (i.e. decay rate $\alpha = 0$), 
               in the present 
               work the stored activation undergoes
               exponential time decay with rate $\alpha$.
  }~\label{fig:neuron} 
  \end{center}
\end{figure}

Recently~\cite{Costa_nrn:2008,Costa_neucomm:2008}, complex neuronal
networks involving simple integrate-and-fire neurons (each neuron is
represented as a node) have been studied with respect to their
transient dynamics.  Figure~\ref{fig:neuron} illustrates the type of
neuronal cell adopted in those works.  The incoming activation,
received through the $n(i)$ dendrites, is integrated and accumulated
in the internal state $S(i)$ until its value exceeds the threshold
$T(i)$, in which case the cell fires, liberating the accumulated
activation between the $m(i)$ outgoing edges (axons).  In the previous
works~\cite{Costa_nrn:2008,Costa_neucomm:2008}, in order to maintain
the total received activation, which was fed through a single selected
neuron, the accumulated activation $S(i)$ was uniformly distributed
among the $m(i)$ outgoing connections, each therefore receiving a
share of $S(i)/k_{out}(i)$, where $k_out(i) = m(i)$ is the out-degree
of node $i$.

Several interesting dynamic features are implied by such a simple
neuronal model.  First, the accumulation of the received activity is
related to the important phenomenon of \emph{facilitation} of firing.
Roughly speaking, the income of a spike into a cell enhances the
probability of its future spiking by occasion of subsequent
activations.  Second, the non-linear element implies the activation to
remain stored until the threshold is reached, which contributes
strongly to constraining the activation locally in the network along
topology and time.  As all neurons are henceforth assumed to have the
same threshold $T=1$ (a biologically reasonable choice), the
distribution of outgoing activation implied by each spiking becomes
imperative in order not to yield one spike at every time step.
Similar effects can be obtained by associating weights smaller or
equal to one to each edge (synaptic weight).  The combination of such
non-linear effects has been observed~\cite{Costa_neucomm:2008} to
contribute decisively for constraining, along a transient period of
time, the activation inside the community which contains the source of
activation.  Such an effect allows the identification of neuronal
communities by considering the transient non-linear dynamics in the
whole network while it is stimulated by sources of activations placed
at each of its neurons.  It has been experimentally verified that the
time it takes for each cell to receive non-zero activation in any of
its dendrites, called the \emph{beginning activation time} of each
cell, seems to be particularly relevant for the identification of the
communities.  Promising results were obtained with respect to two
synthetic (networks including 3 and 4 communities with uniform
connectivity) as well as a real-world network
(\emph{C. elegans}~\cite{Watts_Strogatz:1998}).

However, the previous investigations reported
in~\cite{Costa_neucomm:2008} considered no time decay of the stored
activation $S(i)$, which seems to have been responsible for making the
beginning activation times decisive for the proper identification of
the communities.  In the present work we consider the more biologically
realistic situation involving exponential decays of the activations.
More specifically, at each time step each stored activation is
decreased at a constant rate $\alpha$, i.e.

\begin{equation}
  S^{t+1}(i) = S^{t}(i) - \alpha S^{t}(i)
\end{equation}

where $t$ is the time step and $0 \leq \alpha < 1$.

The net effect of the decay is to generally delay the firing of cells.
Interestingly, such an effect seems to allow proper identification of
the communities also by considering the average activation of the
network along an interval of the transient dynamics, instead of only
the beginning activation time.  This is possibly a consequence of the
fact of the enlarged period of time required to convey the activation
from one community to another, which is enhanced by the decays.  This
possibility is experimentally investigated in the current article by
considering hybrid networks containing four communities of different
types (Erd\H{o}-R\'enyi, Barab\'asi-Albert, Watts-Strogatz as well as
a simple geographical model).  Several combinations of inter and
intra-community intensities of connections are considered.  The
activation is averaged from the beginning of the source operation for
a total of $H$ steps along the transient dynamics.  Then, the
statistical method known as Principal Component Analysis (PCA) is
applied to the activation patterns in order to reduce their
dimensionality, which is optimally obtained by decorrelation of the
activation.  The original communities could be properly detected in
most cases, even for the Barab\'asi-Albert and geographical models.
Combined with the investigations reported
previously~\cite{Costa_neucomm:2008}, the results obtained in the
current work substantiate further the importance of the transient
regime for characterization of modularity regarding both structure and
dynamics in complex systems.  With respect to the specific area of
neuronal networks, the relationships between structured connectivity,
in the form of communities, and the activation and spiking dynamics
provide several implications for synchronization, pattern recognition
and memory.  The proposed methodologies may also prove useful as
practical methods for identification of communities in more general
types of networks.

The current article starts by presenting the basic concepts in complex
neuronal networks, the four adopted theoretical models of complex
networks, and the statistical method of Principal Component Analysis.
The results, discussion, and perspectives for future works are
presented subsequently.

\section{Basic Concepts}

A directed, unweighted network $\Gamma$ can be completely specified in
terms of its \emph{adjacency matrix} $K$.  Each edge extending from
node $i$ to node $j$ is represented $K(j,i) = 1$.  The absence of
connection between nodes $i$ and $j$ implies $K(j,i)=0$.  The nodes
which receive a direct edge from a node $i$ are called the
\emph{immediate neighbors} of $i$.  The \emph{out-degree} of a node $i$ 
is equal to the  number of its immediate neighbors.  

Four theoretical models of complex networks
(e.g.~\cite{Albert_Barab:2002, Newman:2003, Dorogov_Mendes:2002,
Costa_surv:2007}) have been used in order to construct the hybrid
community networks considered in this work: Erd\H{o}s-R\'enyi (ER),
Barab\'asi-Albert (BA), Watts-Strogatz (WS) as well as a simple
geographical type of network (GG).  An Erd\H{o}-R\'enyi network (see
also~\cite{Flory}) can be obtained by establishing connections between
pairs of nodes with constant probability.  The BA networks were
obtained by starting with $m0$ nodes and progressively incorporating
new nodes with $m$ edges, which are attached to the remainder nodes
with probability proportional to their respective degrees.  The WS
structures were obtained by starting with a linear regular network of
suitable degree and subsequently rewiring $10\%$ of its edges.  The
geographical structures are obtained by distributing $N$ nodes along a
two-dimensional space and then connecting each pair of nodes whose
distance does not exceed a given threshold.  Though all these networks
are \emph{undirected}, we obtained the respective directed neuronal
complex networks by considering the incoming on outgoing directions of
each edge as dendrites and axons, respectively.  Therefore, the
so-obtained networks are directed and have in-degree identical to the
out-degree.

The \emph{integrate-and-fire neuron} adopted in this work has been
described and discussed in the Introduction.  The activation and
spiking of all neurons in the network can be represented in terms of
diagrams which are henceforth called \emph{activogram} and
\emph{spikegram}, respectively.  These diagrams are matrices
storing the transient activation or occurrence of spikes for every
node.  In this article, the activation of the network is always
performed by injecting external activation of intensity $1$ at each of
the neurons.  The time it takes for each neuron $i$, from the onset of
the external initiation, to receive the first non-zero input is
henceforth called its respective ~\emph{beginning activation time}
$T_a(i,v)$.  The time it takes for that neuron to produce the first
spike is the~\emph{beginning spiking time} $T_s(i,v)$.

Because the activation patterns obtained with the source in each of
the $N$ neurons involve $N$ measurements, a highly dimensional space
is implied.  As a consequence of the intrinsic correlations between
the activation patterns, it is possible to apply the PCA method to
optimally decorrelated those patterns and yield meaningful 2D and 3D
projections.  Let each of the $N$ observations $v =
\{1, 2,\ldots, N \}$ be characterized by the average activations 
of all nodes as a consequence of the activation source placed at node
$v$.  These measurements can be organized into respective
\emph{feature vectors} $\vec{f_v}$, with elements $f_v(i)$, $i
\in \{1, 2, \ldots, N\}$.  Let the \emph{covariance matrix} between
each pair of measurements $i$ and $j$ be defined as

\begin{equation}
  C(i,j) = \frac{1}{N-1} \sum_{v=1}^{N} (f_v(i) - \mu_i)(f_v(j) - \mu_j)
\end{equation}

where $\mu_i$ is the average of $f_v(i)$ considering all the $N$
observations (i.e. activations).  The eigenvalues of $C$, sorted in
decreasing order, are henceforth represented as $\lambda_i$, $i = 1,
2, \ldots, M$, with respective eigenvectors $\vec{v_i}$.  The matrix
$G$ given in Equation~\ref{eq:PCA}, obtained from the eigenvectors of
the covariance matrix, defines the stochastic linear transformation
known as the \emph{Karhunen-Lo\`eve
Transform}~\cite{Costa_book:2001,Costa_surv:2007}.

\begin{equation} \label{eq:PCA}
  G  =  \left [ \begin{array}{ccc}
              \longleftarrow  &  \vec{v_1}  &  \longrightarrow  \\
              \longleftarrow  &  \vec{v_2}  &  \longrightarrow  \\
              \ldots          &  \ldots     &  \ldots  \\
              \longleftarrow  &  \vec{v_m}  &  \longrightarrow  \\
             \end{array}   \right]  
\end{equation}

with $m=N$.  Because such a transformation optimally decorrelates the
activation patterns, concentrating the variance of the observations
along the first axes (the so-called principal axes or variables), it
is frequently possible to reduce the dimensionality of the
measurements without substantial loss of information by considering 
the above matrix with $m \ll N$. The new, projected measurements
$\vec{g}$, with dimension $m$, can now be straightforwardly obtained
in terms of the following linear transformation

\begin{equation}
  \vec{g} = G \vec{f}.
\end{equation}

\section{Results and Discussion}

\begin{figure*}[htb]
  \begin{center}
  \hspace{1cm} $k =0.1$  \hspace{4cm}  $k =0.5$   
        \hspace{4cm}  $ k = 1$  \\
  $m=2$  
  \includegraphics[width=0.3\linewidth]{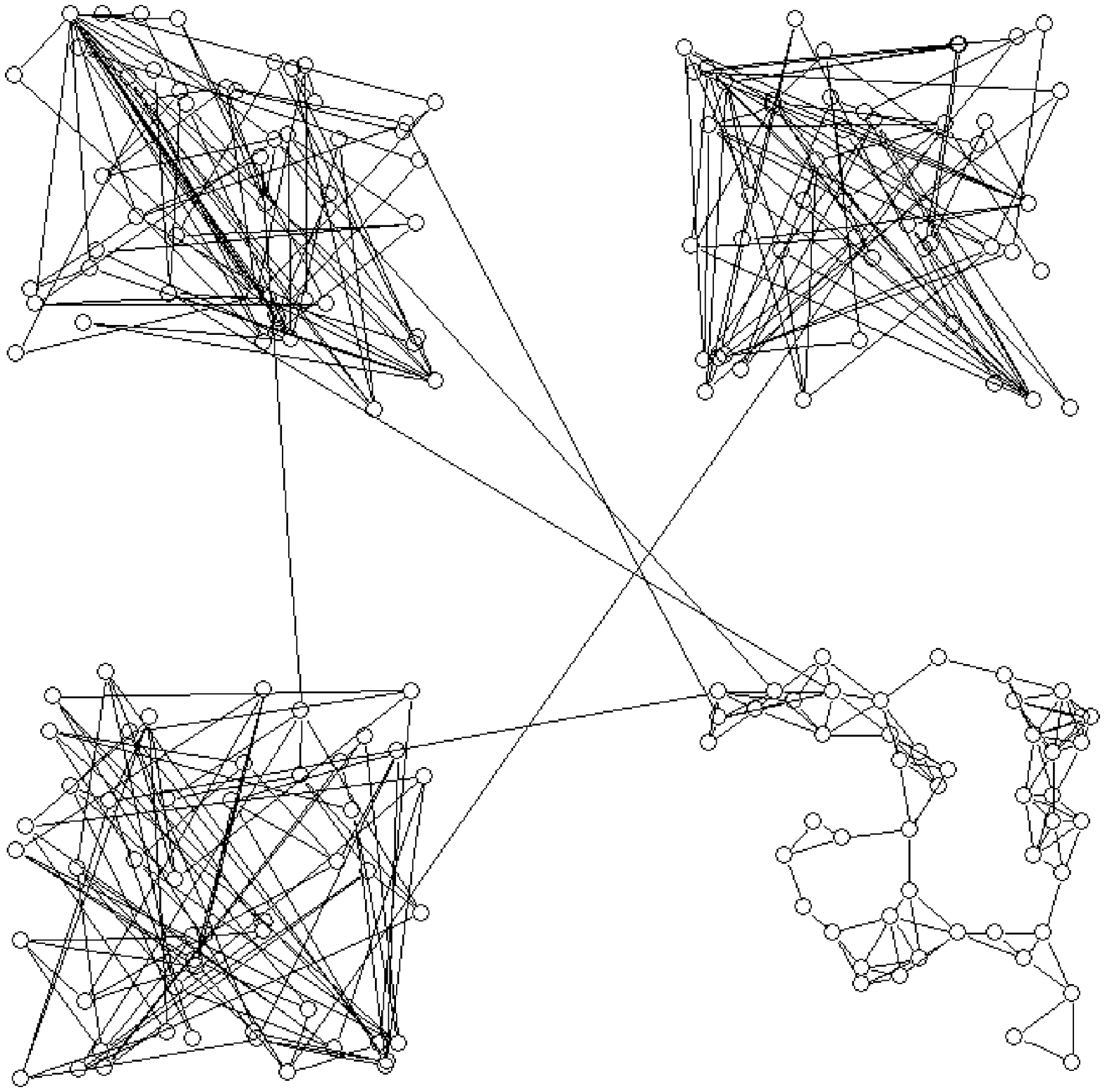} 
  \includegraphics[width=0.3\linewidth]{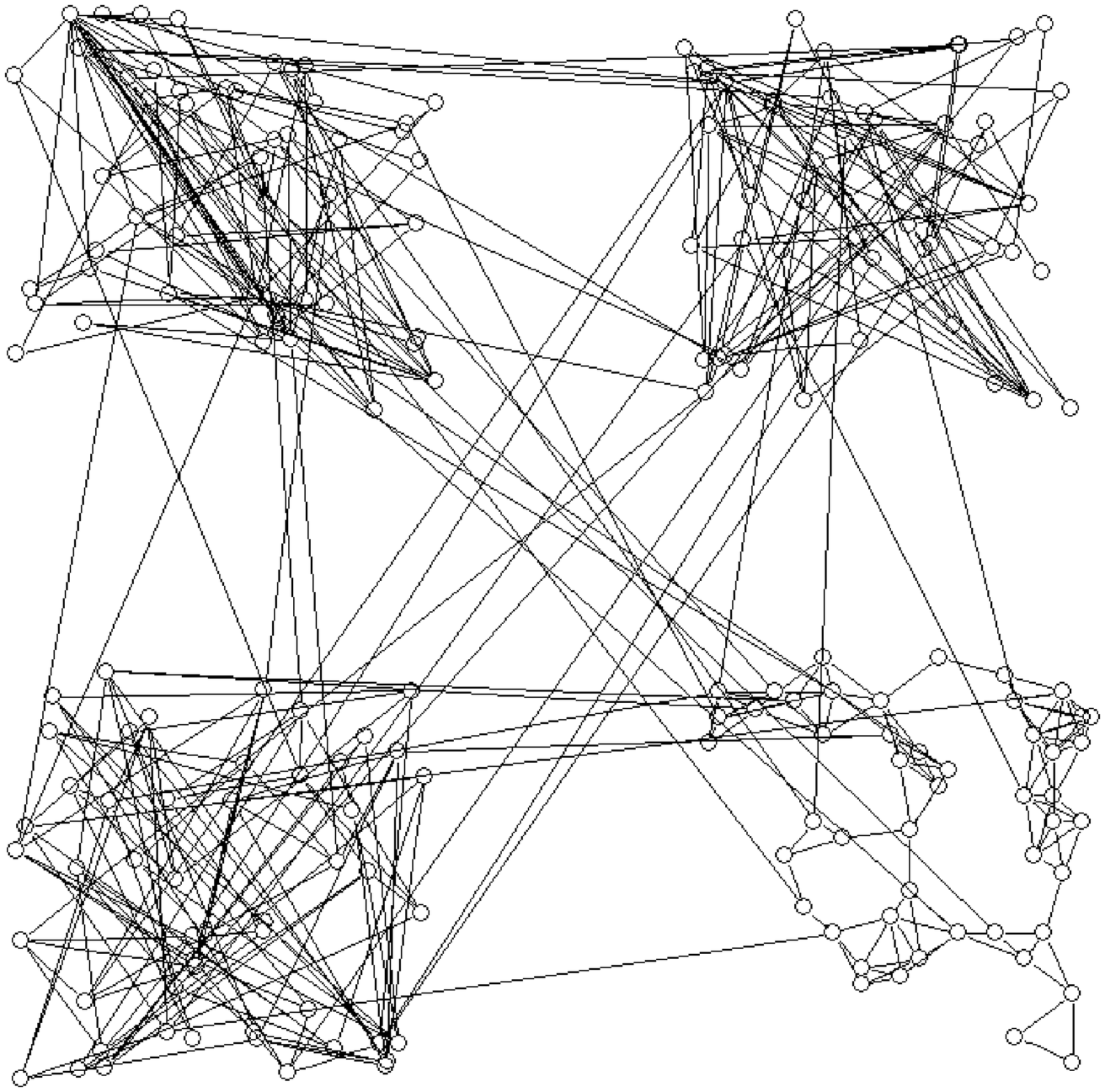} 
  \includegraphics[width=0.3\linewidth]{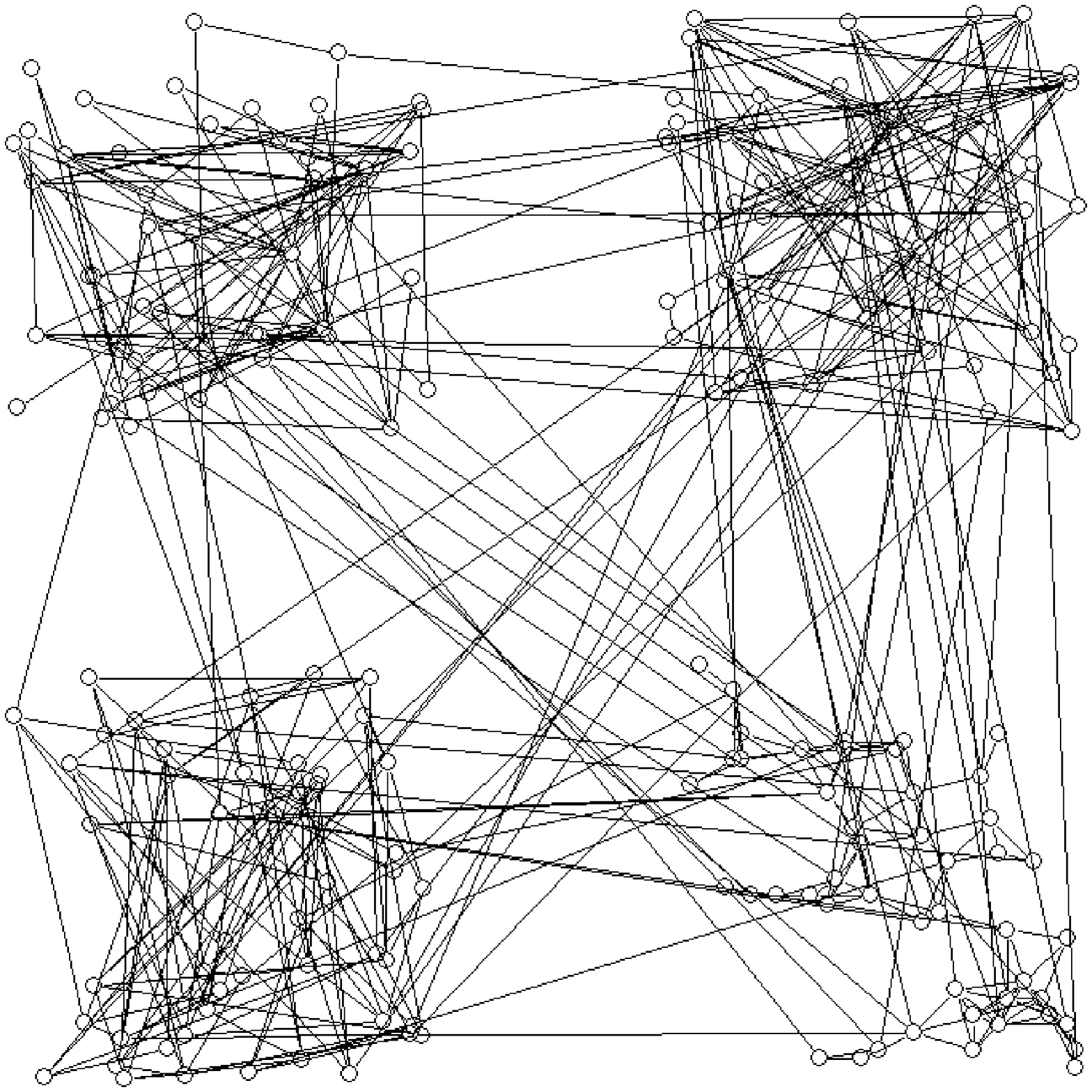} \\ 
  \hspace{1cm} (a)  \hspace{4.7cm}  (b)  \hspace{4.7cm}  (c)  \\
  $m=3$  
  \includegraphics[width=0.3\linewidth]{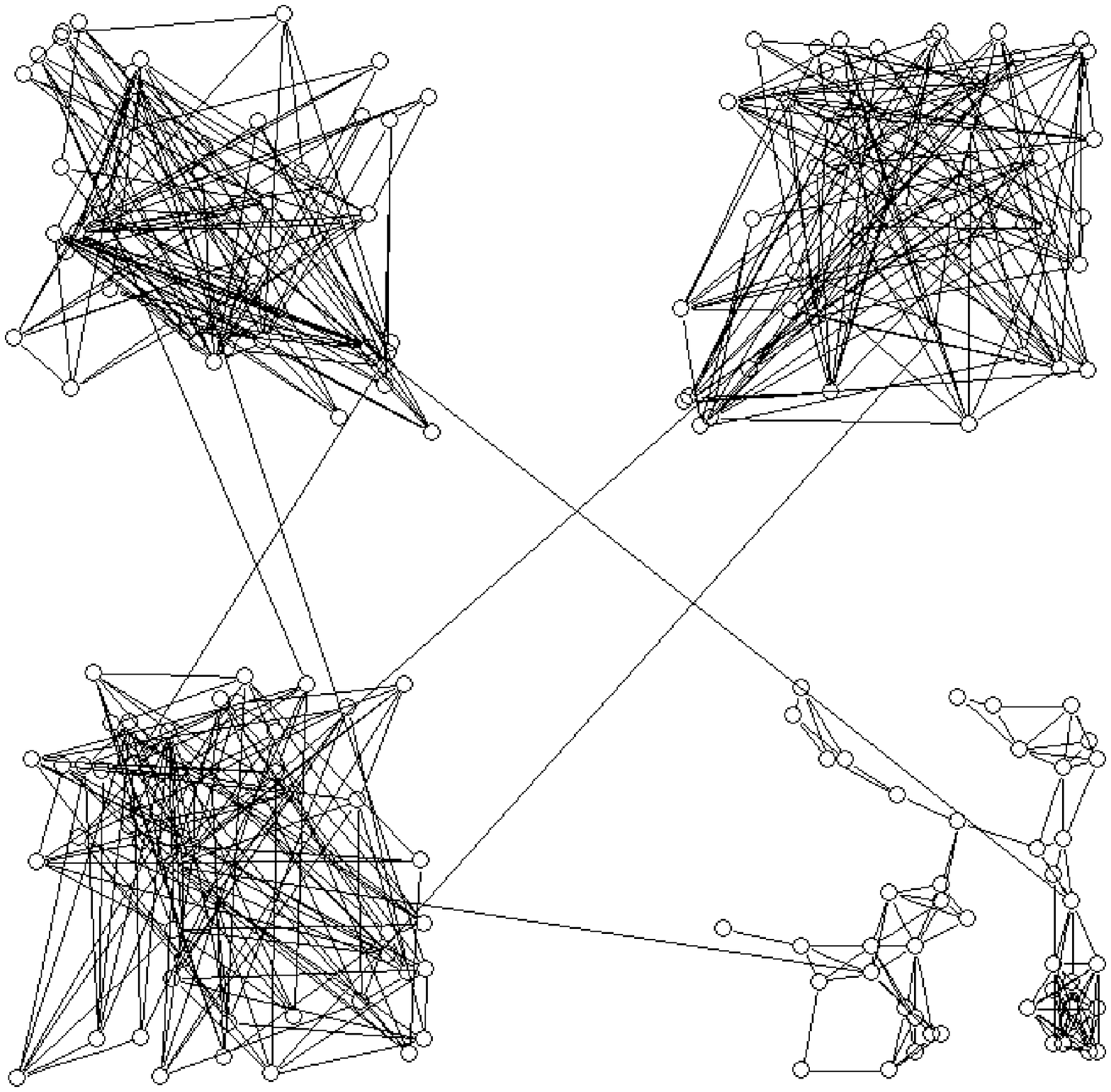} 
  \includegraphics[width=0.3\linewidth]{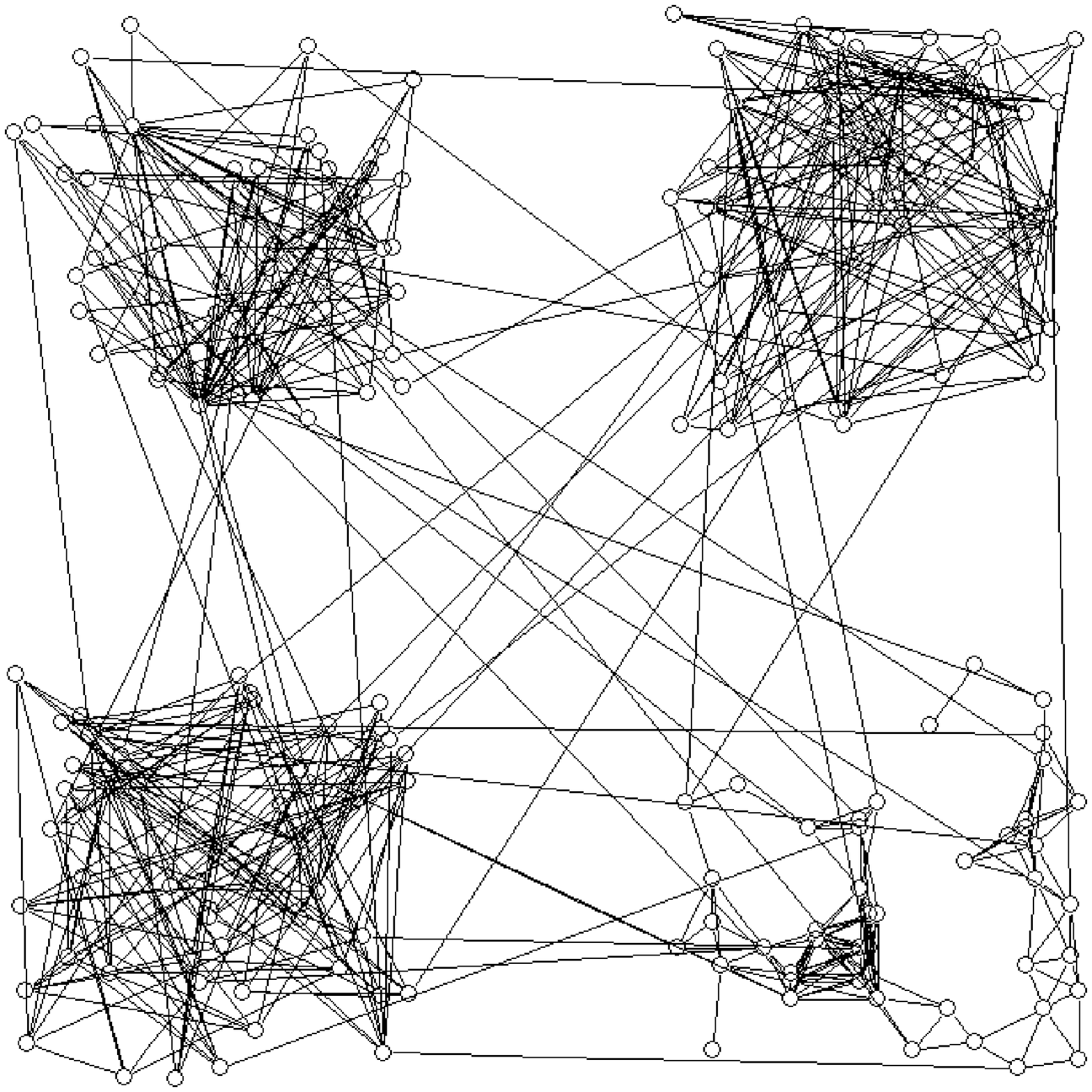} 
  \includegraphics[width=0.3\linewidth]{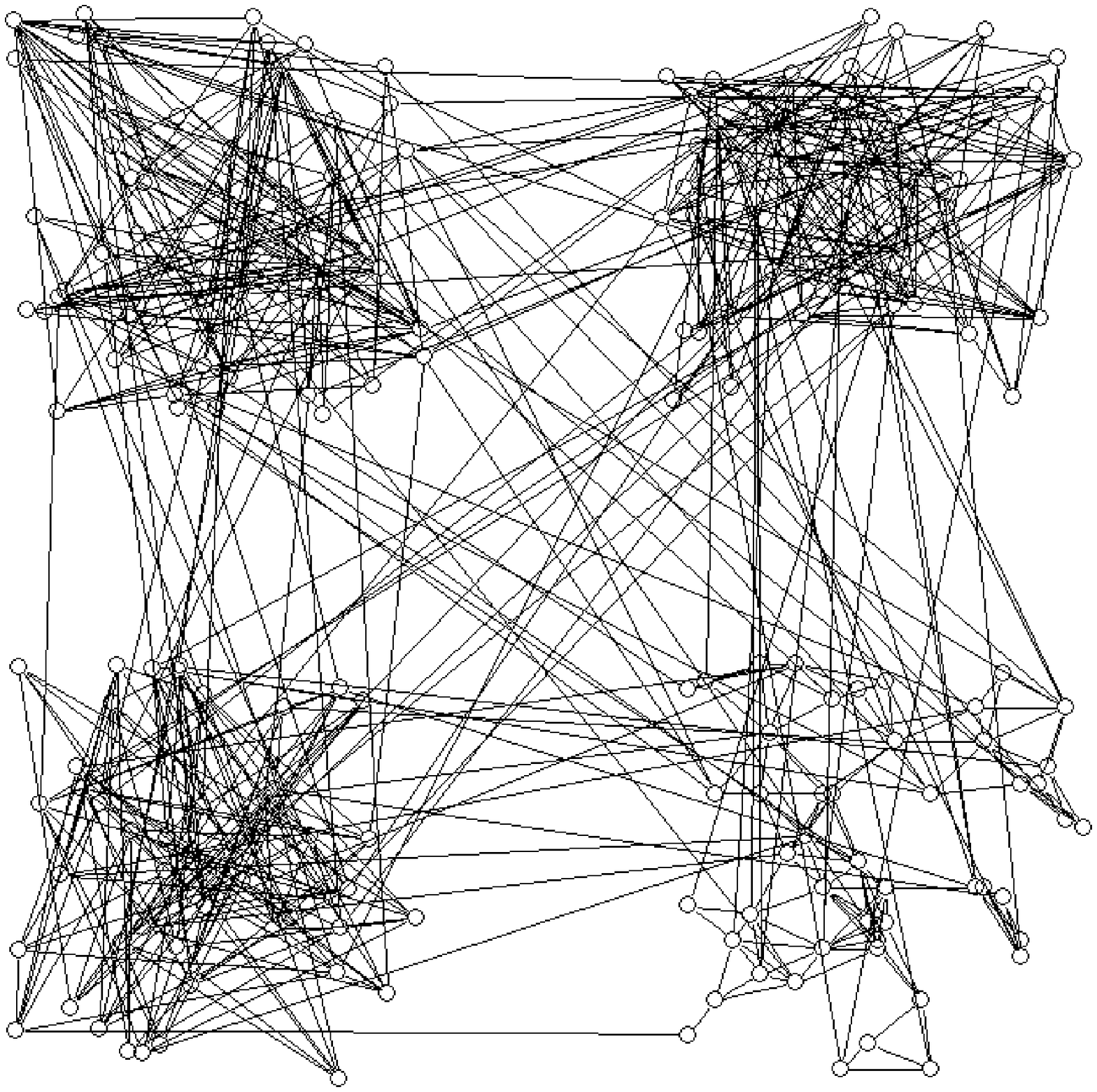} \\ 
  \hspace{1cm} (d)  \hspace{4.7cm}  (e)  \hspace{4.7cm}  (f)  \\
  $m=4$  
  \includegraphics[width=0.3\linewidth]{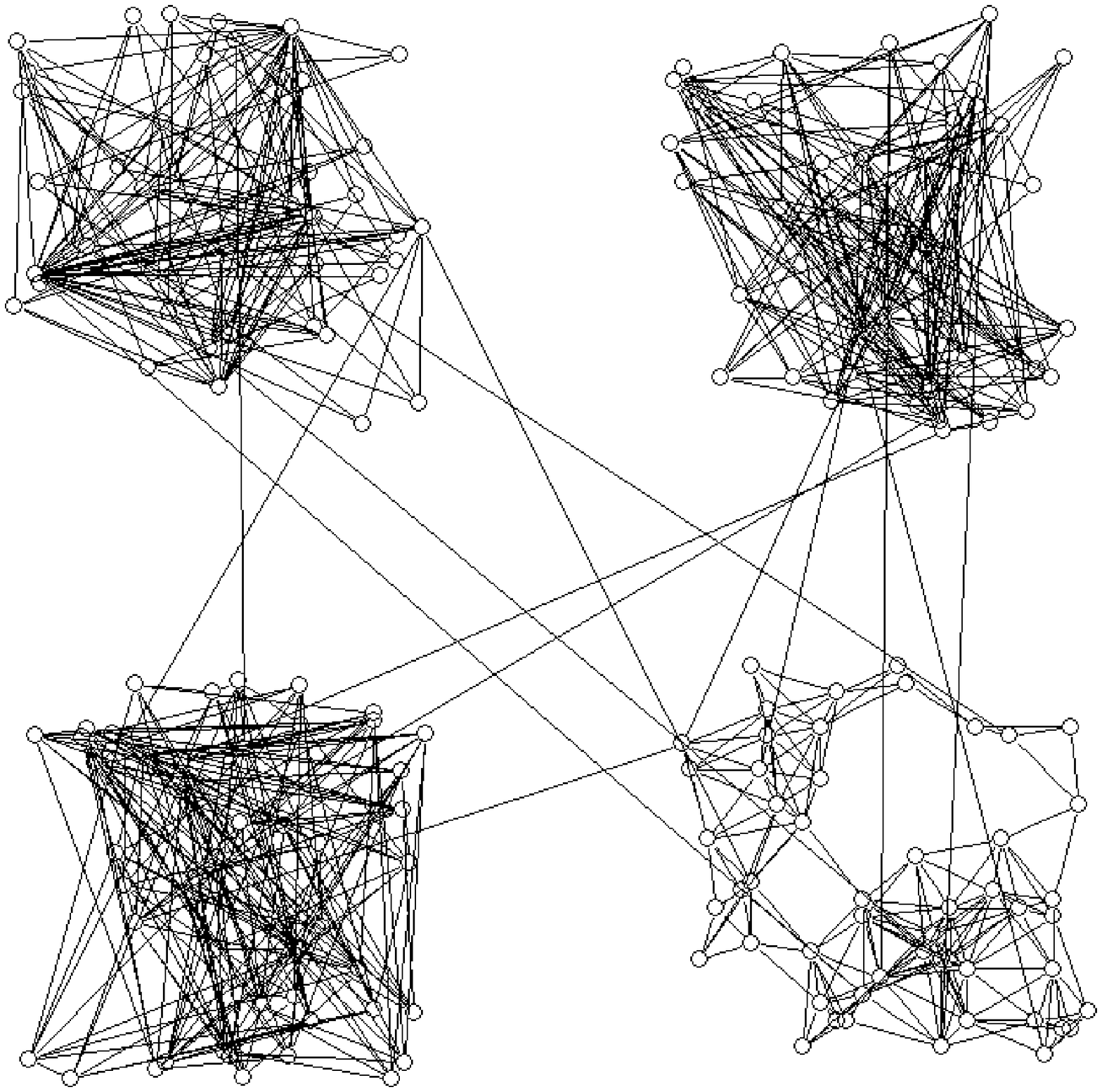} 
  \includegraphics[width=0.3\linewidth]{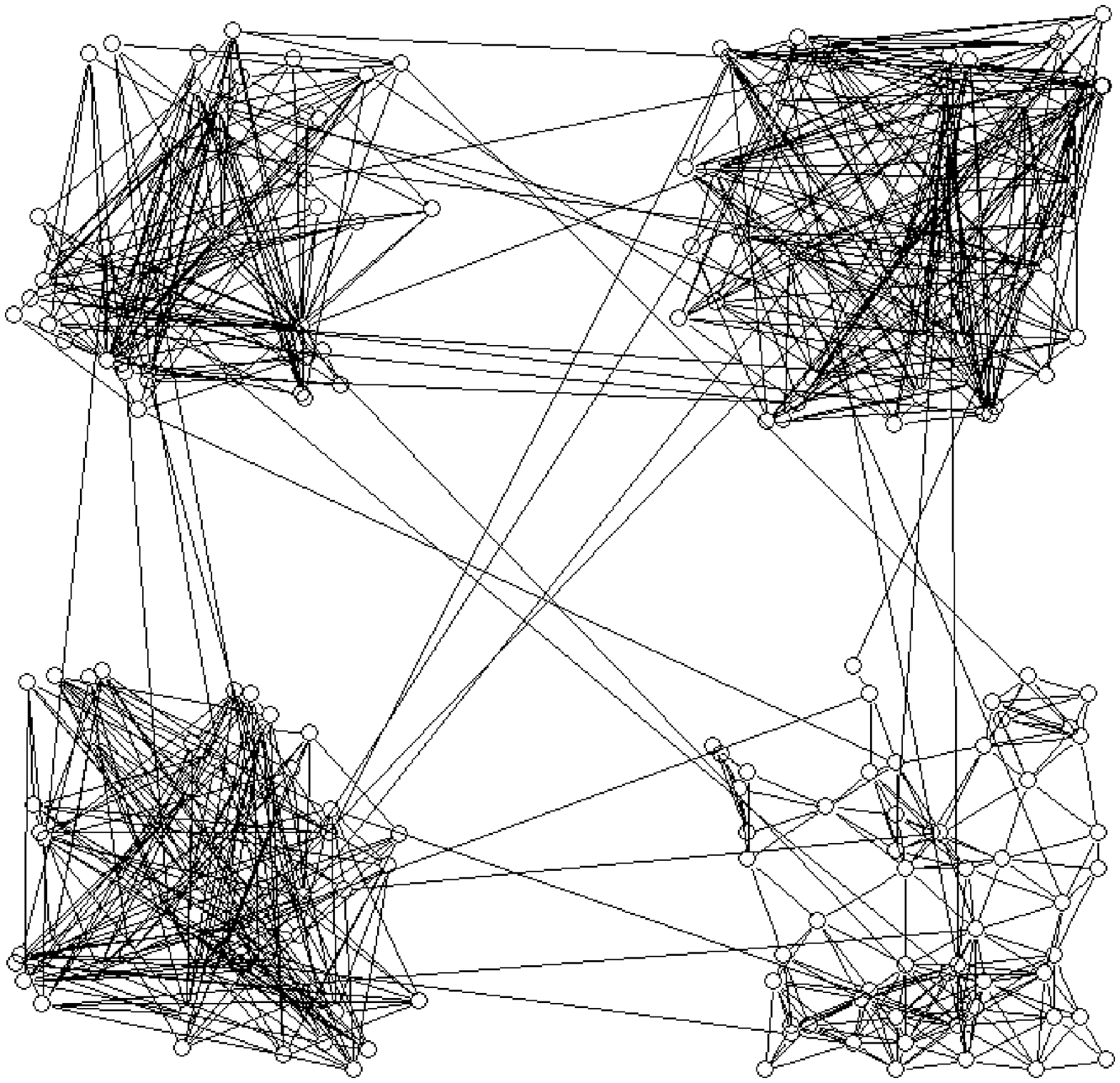} 
  \includegraphics[width=0.3\linewidth]{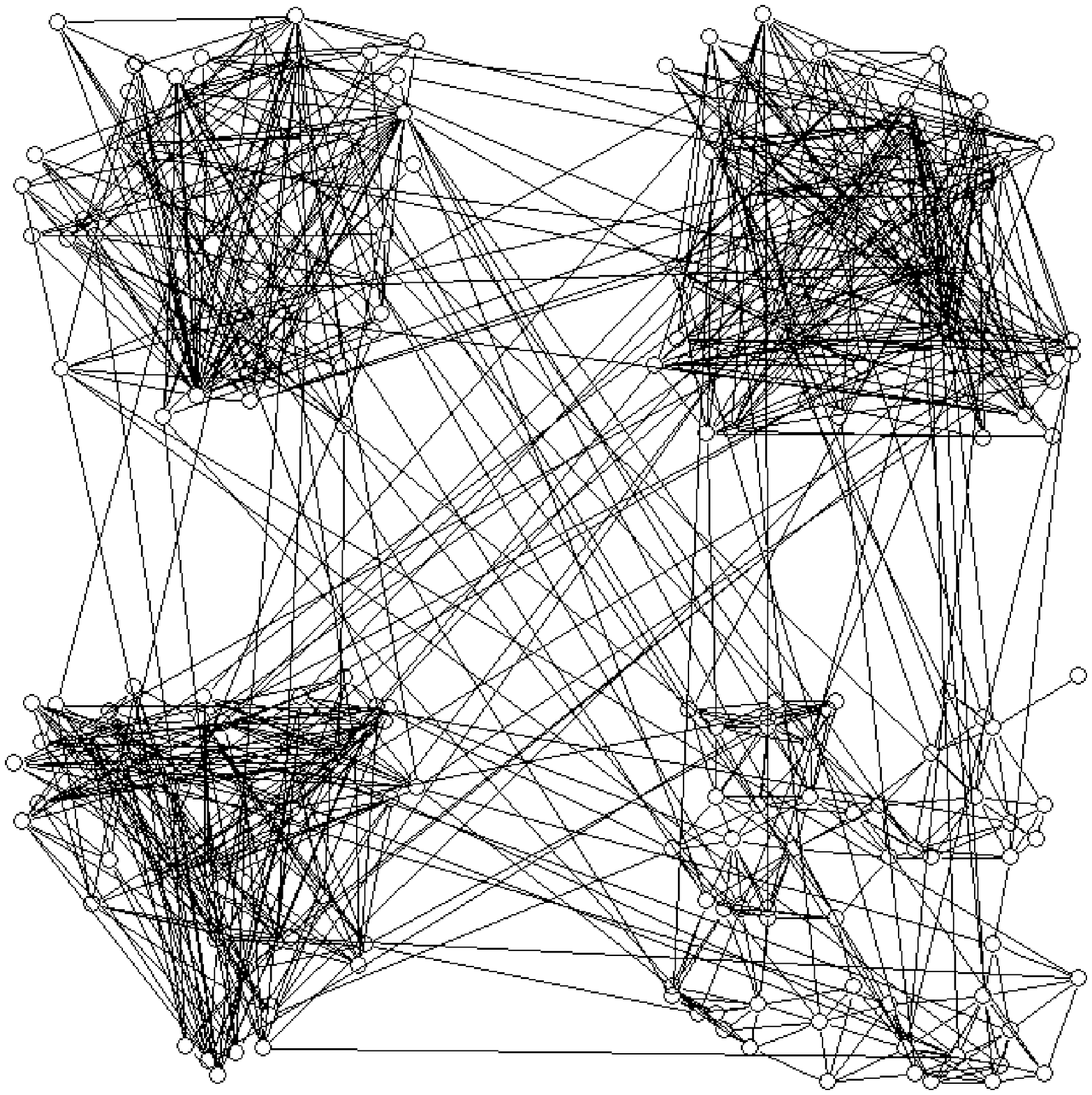} \\ 
  \hspace{1cm} (g)  \hspace{4.7cm}  (h)  \hspace{4.7cm}  (i)  \\
  \end{center}
  \vspace{0.7cm}     legend:     
  \includegraphics[width=0.1\linewidth]{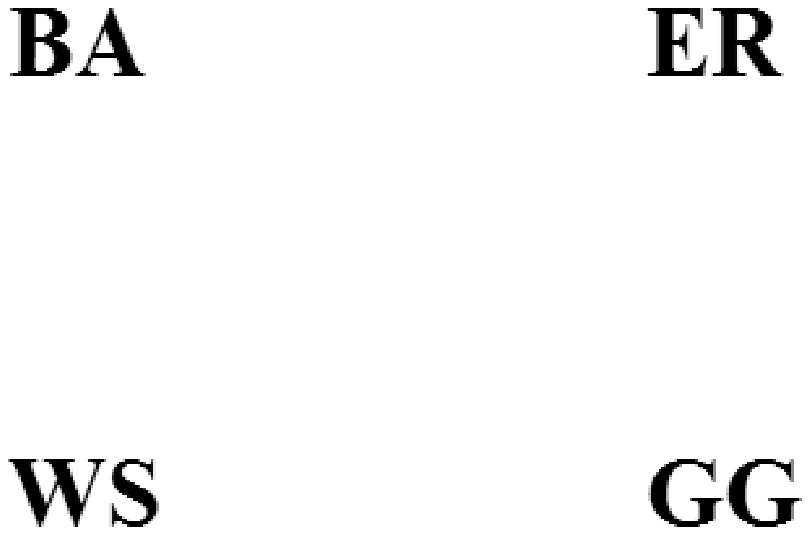} 
  \vspace{0.7cm}  
  \caption{The 9 hybrid networks considered in this work incorporate
                4 communities each, of respective ER, BA, WS and GG 
                types (see legend at the bottom).
  }~\label{fig:nets} 
\end{figure*}

Figure~\ref{fig:nets} illustrates the 9 networks adopted in the
present investigation.  Each of them involves 4 communities, of
respective ER, BA, WS and GG types (see legend at the bottom of the
figure) and approximately 50 nodes each.  The intra-community degrees,
expressed in terms of the BA parameter $m$, increase along the columns
(top to bottom), and the inter-community degrees $k$ increase along
the rows (left to right).  The considered values of intra- and
inter-connectivity are shown in Figure~\ref{fig:nets}.  The same
intra-connectivity degree, defined with respect to the parameter $m$
of the BA model, was adopted for all the 4 communities in each case.
The consideration of hybrid communities involving several network
models is particularly useful for investigating the community
detection methodology with respect to varying connectivity patterns.

\begin{figure*}[htb]
  \vspace{0.3cm} \begin{center}
  \includegraphics[width=0.7\linewidth]{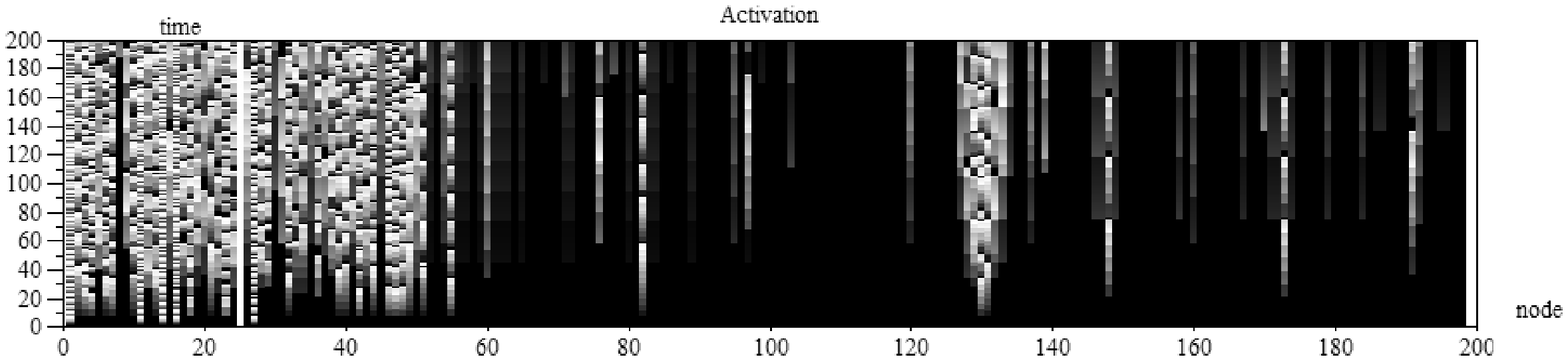}  (a) \\
  \includegraphics[width=0.7\linewidth]{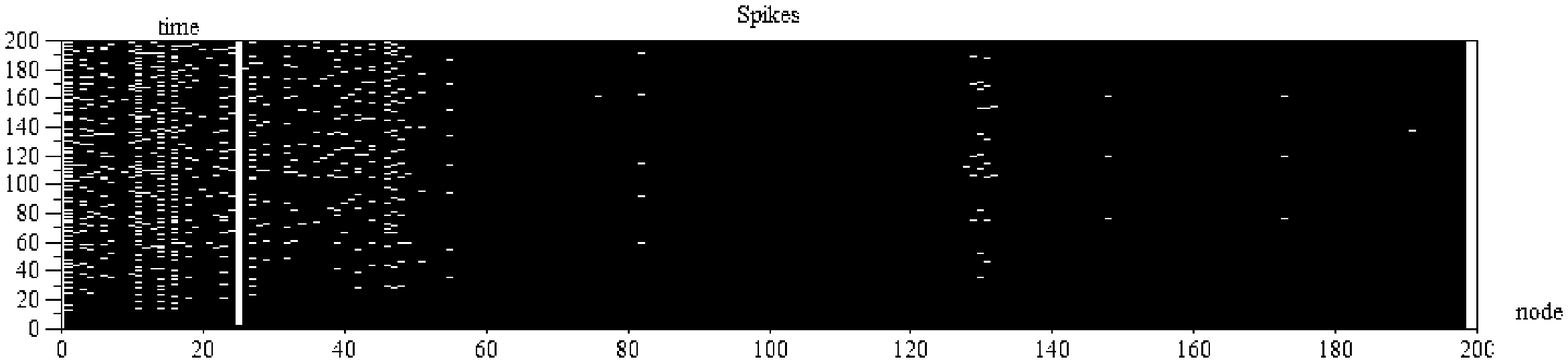}  (b) \\
  \includegraphics[width=0.7\linewidth]{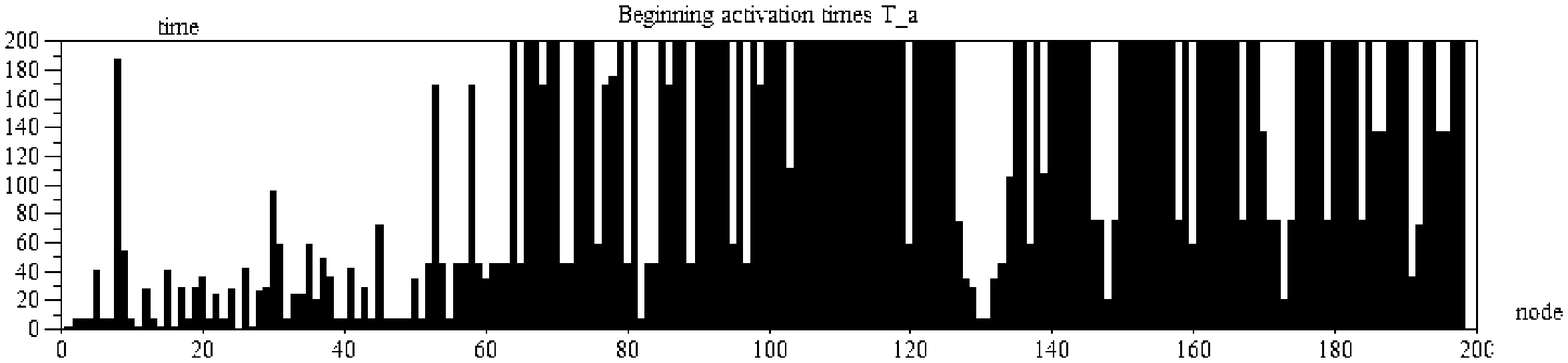}  (c) \\
  \includegraphics[width=0.7\linewidth]{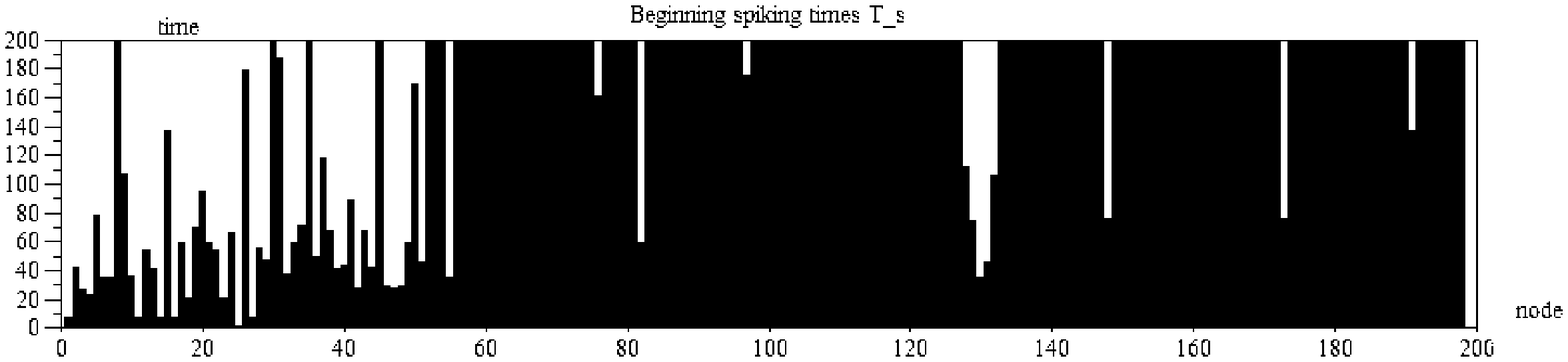}  (d) \\
  \caption{The activogram and spikegram, as well as the diagrams of
              beginning activation times and beginning spiking times
              obtained for the 200 initial time steps with activation 
              source at node 25 for the complex network with $m=3$, 
              $k = 0.5$ (Figure~\ref{fig:nets}e) and $\alpha = 0.02$.
          }~\label{fig:grams} 
  \end{center}
\end{figure*}

Each network was searched for community structure by placing the
activation source (with intensity 1) at each of its neurons and
simulating the respective activation and spiking along the initial
$H=200$ steps of the transient dynamics.  Three whole set of
simulations where performed by considering respective decay rates
$\alpha$ equal to 0.02 and 0.5.  Figure~\ref{fig:grams} shows the
activogram and spikegram, as well as the diagrams of beginning
activation times and beginning spiking times for the network with
$m=3$, $k = 0.2$ and $\alpha = 0.02$.

The constant activation fed through neuron 25 is clearly identified as
the white column in the activogram and spikegram.  Because of the more
intense interconnectivity between the nodes in the community to which
this neuron belongs (ER, in this case) the propagation of activation
and spikes tend to occur first inside this community, being propagated
to the other communities only later.  One exception are the neurons
around node 125, which belong to the WS community.  Because neuron 125
is connected to the ER community with particular intensity (more than
one edge), it receives considerable activation sooner, leading to a
progressive spreading of activation within the WS community.  However,
because this effect is not verified for most of the other nodes of the
ER community, it tends to become less relevant in the subsequent
decorrelation projection implemented by the PCA.  In addition, except
for a few other cells, the nodes which are inside the same community
tend to receive activation relatively soon, as illustrated in the
respective diagram of beginning activation times.  A less regularly
simultaneous activation is obtained for the spikes in the respective
beginning spiking times diagram.  The incorporation of suitable (not
too large) values of decay seems to promote more stable activation
patterns for the source placed at neurons of a same community as a
consequence of further constraints on the dispersion of the
activation.  In this work we consider for the community identification
the patterns of activation obtained by integrating the activation from
time 0 to $H=200$.  Observe that the parameter $H$ has important
implications for the computational cost, in the sense that the larger
its value, the larger the number of computations.

\begin{figure*}[htb]
  \vspace{0.3cm} 
  \begin{center}
  \includegraphics[width=0.9\linewidth]{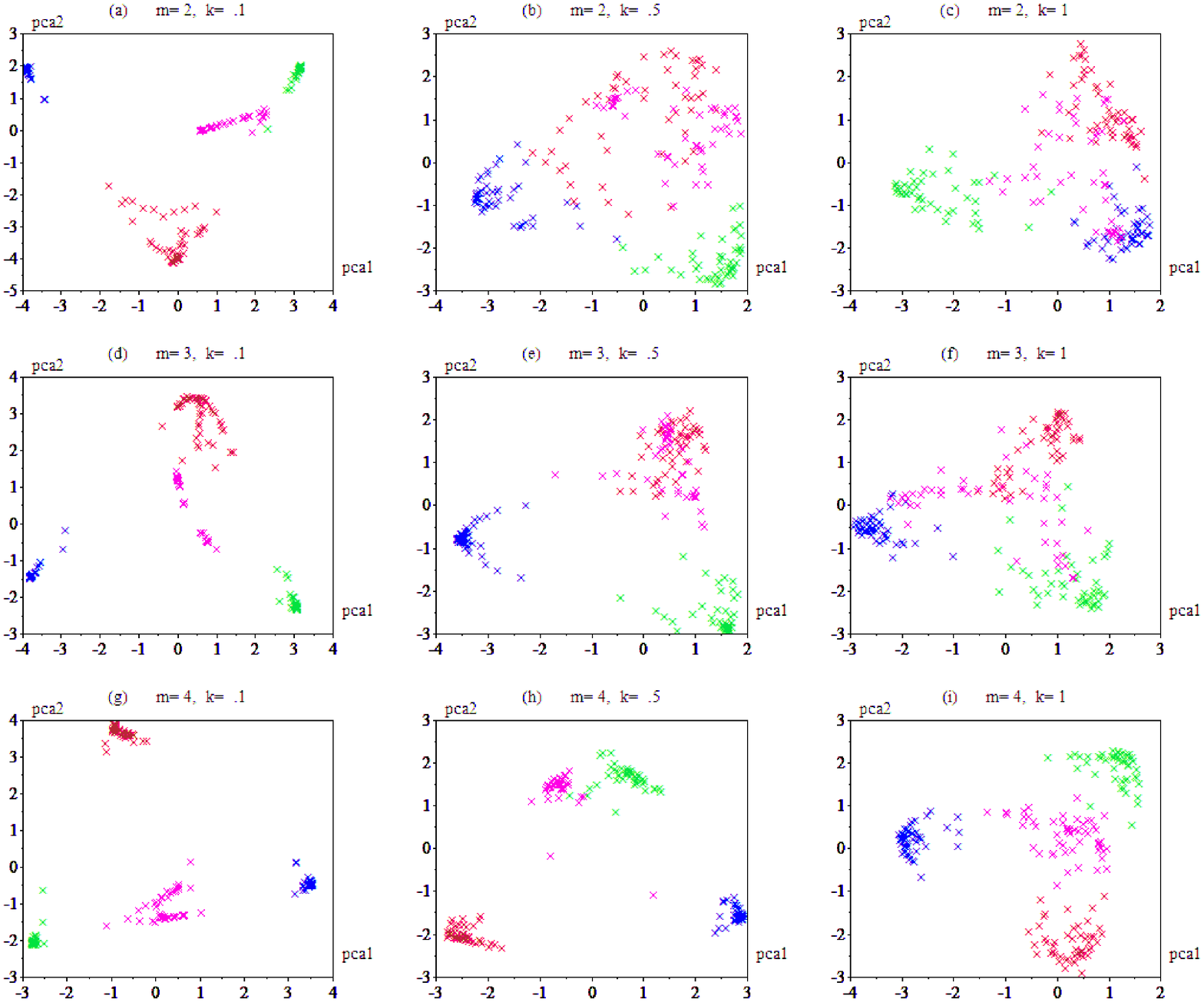} 
  \caption{The clusters obtained by considering the first and
              second PCA variables for $\alpha = 0.02$.
          }~\label{fig:pcas_0p02} 
  \end{center}
\end{figure*}

\begin{figure*}[htb]
  \vspace{0.3cm} 
  \begin{center}
  \includegraphics[width=0.9\linewidth]{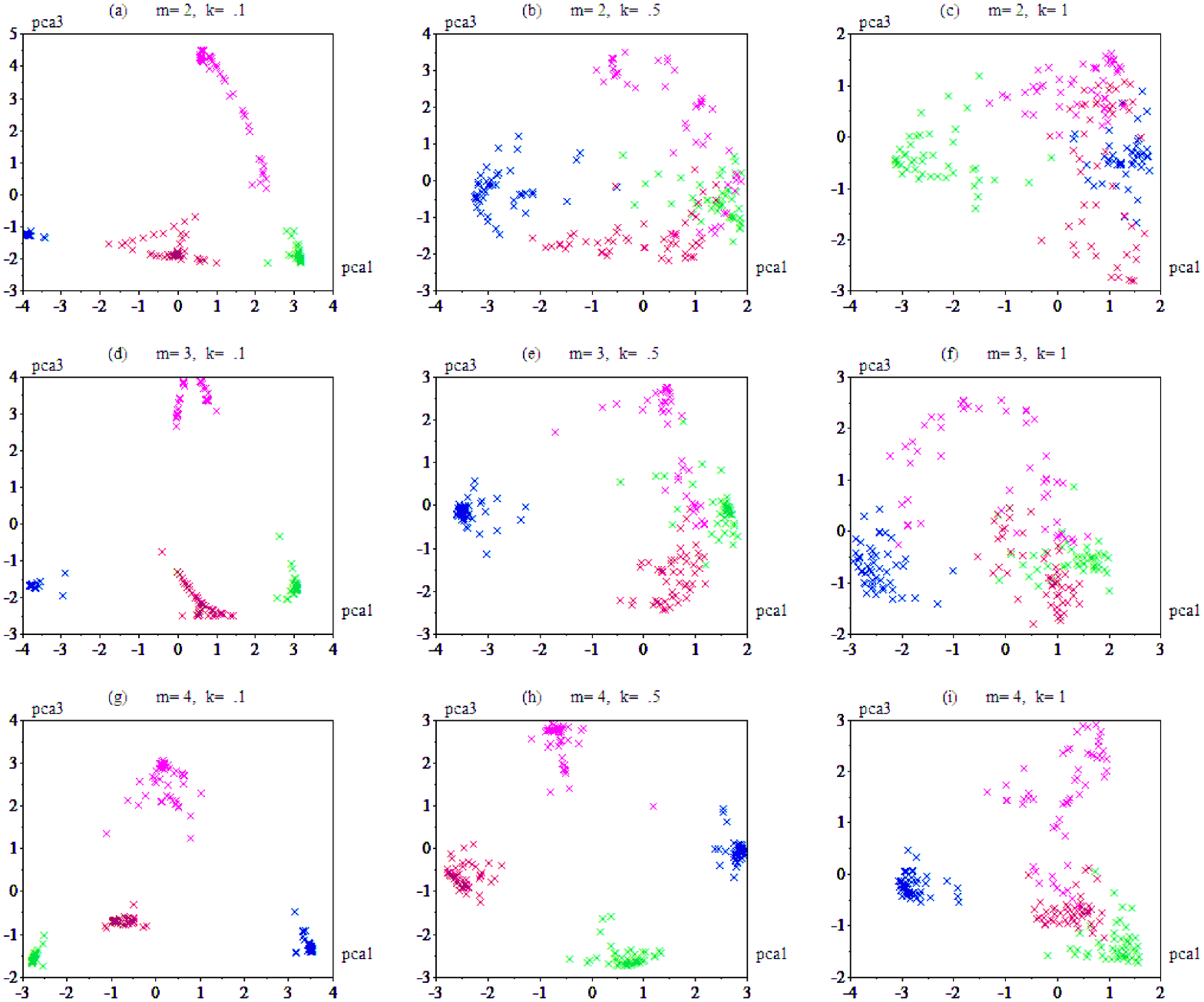} 
  \caption{The clusters obtained by considering the first and
              third PCA variables for $\alpha = 0.02$.  
           }~\label{fig:pcas_0p02_z} 
  \end{center}
\end{figure*}

Figure~\ref{fig:pcas_0p02} depicts the clusters obtained in the
two-dimensional space defined by the first two PCA variables
considering the average activation patterns and decay $\alpha = 0.02$.
Figure~\ref{fig:pcas_0p02_z} shows the respective scatterplots
obtained for the first and third PCA variables. Therefore, it is
possible to have a clear idea of the 3D PCA space by considering these
two images.  In these figures, as well as all the other subsequent
ones, the original communities are identified by respective colors: ER
in blue; BA in green; WS in red and GG in magenta.  In most cases,
especially for low ratios $k / m$, the original communities were
mapped into well-defined respective clusters in the PCA space.  For
instance, in the case $m=2$ and $k=0.1$, we have dense clusters
obtained for the ER and BA communities.  Because of their intrinsic
nature, the WS and GG models tended to yield larger dispersions in
most of the cases considered in this work.  Yet, they are
well-separated, as it can be verified by considering both the $pca1
\times pca2$ (Figure~\ref{fig:pcas_0p02}) and $pca1 \times pca3$
(Figure~\ref{fig:pcas_0p02_z}) diagrams.  Except for the WS case, the
other 3 communities still tended to map to reasonably well-defined
local regions in the PCA projections for larger values of
intercommunity connection (i.e. $k = 0.5$ and $1$).  The separation
between the community clusters tended to increase substantially from
top to down along each column in Figures~\ref{fig:pcas_0p02}
and~\ref{fig:pcas_0p02_z} as a consequence of the increase of the
intra-community connectivity relatively to the inter-community
density of connections.

\begin{figure*}[htb]
  \vspace{0.3cm} 
  \begin{center}
  \includegraphics[width=0.9\linewidth]{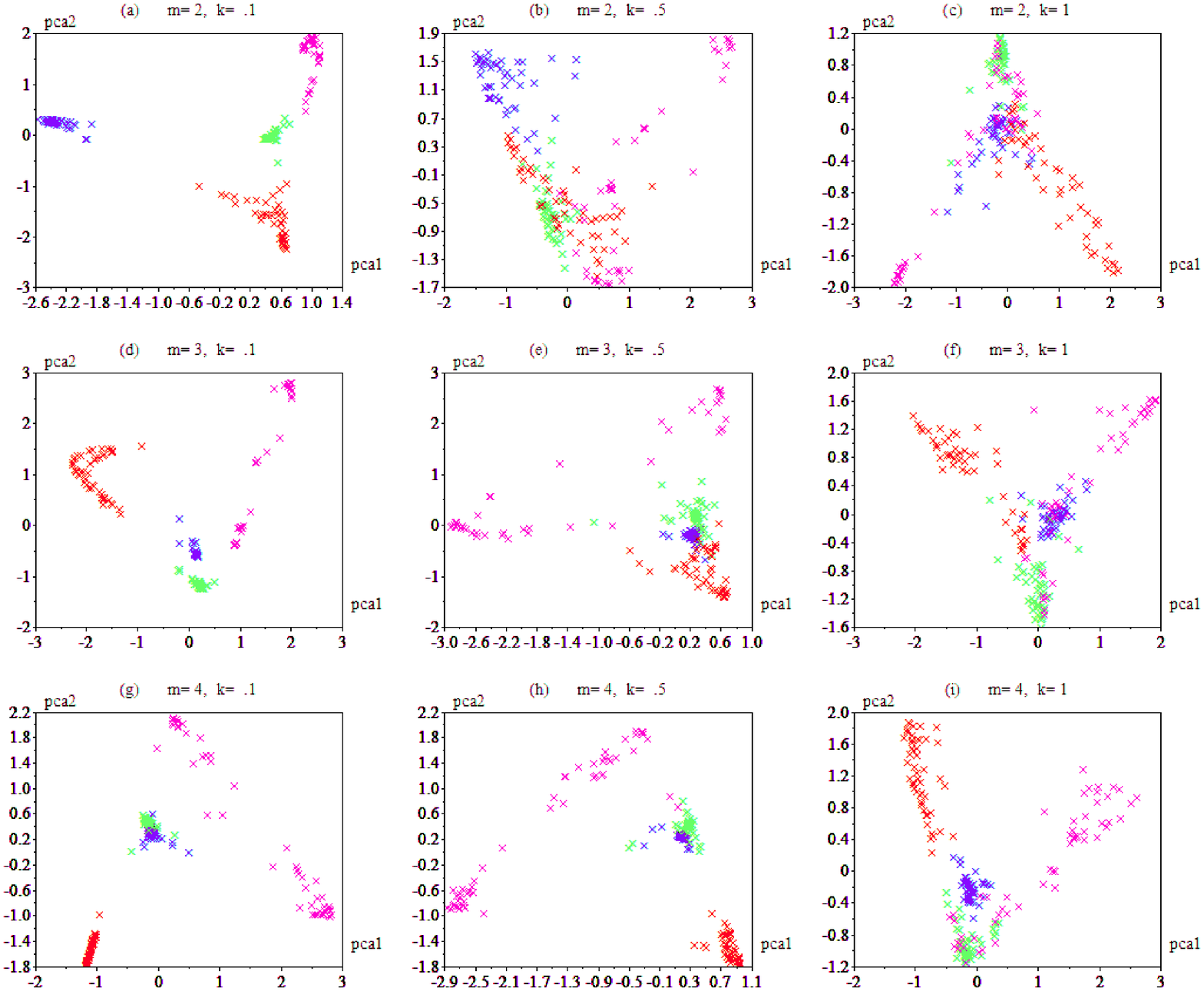} 
  \caption{The clusters obtained by considering the first and
              second PCA variables for $\alpha = 0.5$. 
  }~\label{fig:pcas_0p5} 
  \end{center}
\end{figure*}

\begin{figure*}[htb]
  \vspace{0.3cm} 
  \begin{center}
  \includegraphics[width=0.9\linewidth]{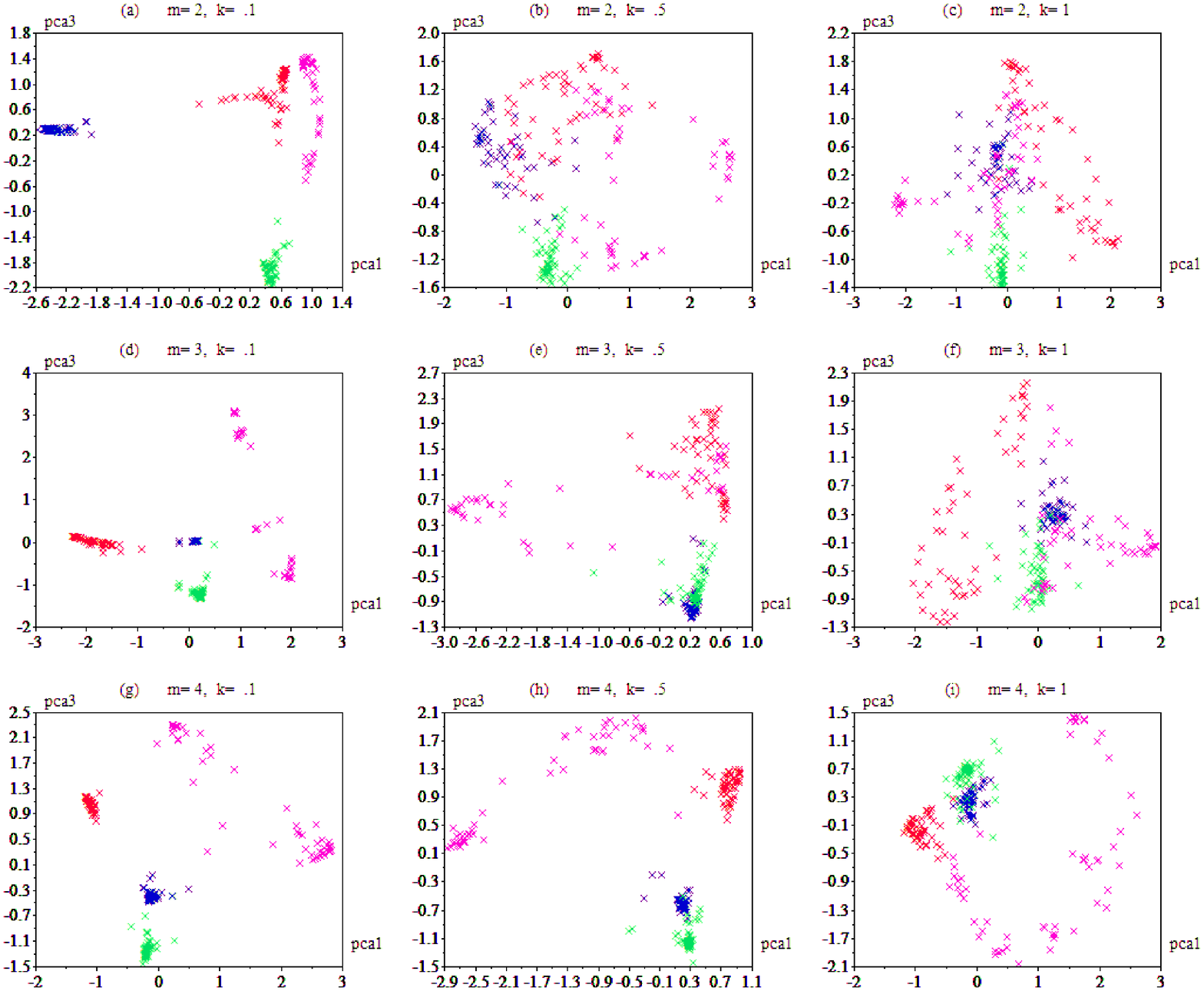} 
  \caption{The clusters obtained by considering the first and
              third PCA variables for $\alpha = 0.5$. 
  }~\label{fig:pcas_0p5_z} 
  \end{center}
\end{figure*}

The PCA scatterplots obtained by considering more intense decay
(i.e. $\alpha = 0.5$) are shown in Figures~\ref{fig:pcas_0p5}
and~\ref{fig:pcas_0p5_z} respectively to the $pca1 \times pca2$ and
$pca1 \times pca3$ projections.  Less separated clusters have been
obtained in most cases, with intense overlap between communities.
However, the nodes belonging to the original communities still tended
to be mapped to nearby positions in the scatterplots.  Such a decrease
in the community identification is a direct consequence of the fact
that more intense decays tended to produce less stable activation
patterns for the activation source placed at different nodes.  In
addition, the consideration of more intense decay also would imply in
averaging the activations along a longer period of time, demanding
additional computations.

\section{Concluding Remarks}

In continuation to recent previous
investigations~\cite{Costa_neucomm:2008}, the relationship between
topological and dynamical modularity during the transient period of
activation of non-linear integrate-and-fire complex neuronal networks
has been explored further, with respect to the consideration of
average activation patterns as resources for community identification.
By increasing the latent period in which the activation has to
increase until firing is reached in the neuronal cells, relatively
small non-zero decays of the accumulated activation tended to allow
the subsequent identification of the topological communities as
clusters appearing in scatterplots obtained by optimally decorrelated
PCA projections.  

Such a phenomenon has been experimentally investigated by considering
several hybrid networks, with communities of different types, and
varying ratios of intra- to intercommunity connectivity.  In most
cases the original communities were mapped onto adjacent sets of
points (clusters) which were often well-defined and delimitated.  As
in~\cite{Costa_neucomm:2008}, the nodes at the interfaces between the
obtained clusters tended to correspond to those nodes implementing the
intercommunity connectionsl.  The ER and BA modules tended to produce
more concentrated clusters, with the WS and GG communities often
yielding scattered, but still separated, distributions in the PCA
diagrams.  The discrimination between the communities was undermined
when a substantially large decay was considered.  As observed
previously~\cite{Costa_neucomm:2008}, the transient confinement of the
activation inside each community seems to be related to an abrupt
pattern of activation observed for several models of complex
networks~\cite{Costa_nrn:2008}.  As a matter of fact, the WS and GG
models had indeed been observed~\cite{Costa_nrn:2008} to produce less
abrupt activations.  The situations involving higher values of
$\alpha$ would also imply substantially higher computational cost
required for the simulation of the activation along longer transient
periods.  Nevertheless, it seems to follow from the currently reported
results that relatively small non-zero decay of the accumulated
activation, to a certain extent, emphasizes the transient confinement
of activation inside the communities.

These findings substantiate the importance of transient dynamics for
the characterization and analysis of non-linear complex systems.
Furthermore, the relationship between modular interconnectivity and
nearly simultaneous activation of communities has several implications
for biological and computational neuroscience.  In particular, such a
relationship can be intrinsically related to recognition of patterns
and associative memory.  For instance, the nearly simultaneous
activation of the communities could play an important role in
reconstructing larger patterns (communities) from incomplete
presentations.  Because of the temporal dynamics of nervous systems,
where problems have to be solved by functional modules of neurons
along a given period of time (e.g.~\cite{Zeki:1999, Hubel:2005}), it
is possible that the phenomenon of simultaneous activation within
communities plays an important general role in neuronal organization
and functionality.  The neuronal community identification methodology
is also promising for the identification of functional modules in the
cortex or neuronal subsystems because of its intrinsic compatibility
with the non-linear dynamics performed in those systems.

Several are the future works implied by the results and methods
reported in the current article.  First, it would be important to
perform more objective investigations of the discriminability between
the PCA clusters, for instance by using the intra- and inter-class
scatterings (e.g.~\cite{Costa_book:2001}) and comparing the results
obtained for the synthetic hybrid communities with those yielded by
canonical analysis (e.g.~\cite{Costa_surv:2007,McLachlan:1998}).  It
would also be necessary to consider larger ensemble of networks in
order to reach more general and definitive conclusions regarding the
effect of the few parameters involved (i.e. $\alpha$ and $H$), as well
as concerining possible finite-size and scaling effects.  Valuable
insights about the influence of the connectivity on the transient
non-linear dynamics of the complex neuronal networks considered in
this work can be potentially achieved by applying the systematic
approach of superedges~\cite{Costa_superedges:2008}.  Of particular
interest are further investigations aimed at the characterization of
the abrupt activations verified for several complex network models.
This phenomenon, which is possibly associated to phase transition
and/or self-organized criticality, seems to lie at the heart of the
confinement of the activation inside the communities during the
transient activation.

\begin{acknowledgments}
Luciano da F. Costa thanks CNPq (308231/03-1) and FAPESP (05/00587-5)
for sponsorship.
\end{acknowledgments}

\bibliography{neudec}
\end{document}